\documentclass[aip,preprint]{revtex4-1}%linenumbers
\usepackage{graphicx}% Include figure files

\draft % marks overfull lines with a black rule on the right

\begin{document}

\title{Quadratic scaling path integral molecular dynamics for fictitious identical particles and its application to fermion systems}

%\title{Quadratic scaling path integral molecular dynamics for fictitious identical particles}

\author{Yunuo Xiong}
\email{xiongyunuo@hbpu.edu.cn}
\affiliation{Center for Fundamental Physics and School of Mathematics and Physics, Hubei Polytechnic University, Huangshi 435003, China}

\author{Shujuan Liu}
\affiliation{Center for Fundamental Physics and School of Mathematics and Physics, Hubei Polytechnic University, Huangshi 435003, China}

\author{Hongwei Xiong}
\email{xionghongwei@hbpu.edu.cn}

\affiliation{Center for Fundamental Physics and School of Mathematics and Physics, Hubei Polytechnic University, Huangshi 435003, China}

\date{\today}

\begin{abstract}

Recently, fictitious identical particles have provided a promising way to overcome the fermion sign problem and have been used in path integral Monte Carlo (PIMC) to accurately simulate warm dense matter with up to 1000 electrons (T. Dornheim et al., arXiv:2311.08098 (2023)). The inclusion of fictitious identical particles in  path integral molecular dynamics (PIMD) can provide another way to simulate fermion systems. In a recent paper (J. Chem. Phys. 159, 154107 (2023)), Feldman and Hirshberg improved the recursive formula for PIMD of N identical bosons, significantly reducing the computational complexity from $O(PN^3)$ to $O(N^2+PN)$. In this paper, we extend this latest recursive formula for bosons to PIMD of fictitious identical particles to improve the efficiency of simulating fermion systems. We also provide the virial estimator for calculating energy by using the recursive technique. As an example, we use the quadratic scaling PIMD for fictitious identical particles to study the simulation of hundreds of fermions in a two-dimensional periodic potential, in the hope of providing a simulation tool for two-dimensional Fermi-Hubbard model and other strongly correlated fermion systems, such as the simulation of ultracold fermionic gases in optical lattices.
\end{abstract}

\maketitle

\section{Introduction}

For a quantum many-body system with $N$ particles, the numerical expression of the wave function $\Psi(\textbf r_1,\cdots,\textbf r_N)$ increases exponentially with the increase of the number of particles. For example, in three-dimensional space, if each dimension is divided into $M$ parts for approximation, then the numerical expression of the wave function of the quantum many-body system requires $M^{3N}$ orders of magnitude of complex numbers \cite{CeperleyBook}. When we use numerical simulation to study the thermodynamic properties of quantum many-body systems, the path integral formalism can well solve the difficulty of being unable to numerically express the wave function of quantum many-body systems. In the path integral formalism \cite{Feynman,Tuckerman}, the treatment of the partition function involves high-dimensional integration, which can be handled by Monte Carlo or molecular dynamics, resulting in  path integral Monte Carlo (PIMC) and  path integral molecular dynamics (PIMD)\cite{barker,Morita,CeperleyRMP} for the exact numerical simulation of the thermodynamic properties of quantum many-body systems.

PIMC and PIMD have successfully solved the problem of exact numerical simulation of the thermodynamic properties of large-scale bosonic systems \cite{CeperleyRMP,Burov1,Burov1b}. Unfortunately, for PIMC/PIMD, when simulating fermions, the fermion sign problem \cite{ceperley,Alex,troyer,Dornheim,WDM} poses a serious obstacle, and the computational cost increases exponentially with the number of fermions or the decrease in temperature. Recently, we used the idea of fictitious identical particles \cite{XiongFSP,Xiong-xi} to try to overcome the fermion sign problem, and found that it is indeed promising to accurately simulate the thermodynamic properties of many important fermionic systems from zero temperature to high temperature, and the computational complexity is no different from the case of bosons. Afterwards, the idea of fictitious identical particles for overcoming the fermion sign problem was systematically verified and confirmed by Dornheim et al. \cite{Dornheim1}, and after extending the applicability of fictitious identical particles, they highly accurately simulated the static structure factor and static linear density response function of warm dense matter with one thousand fermions \cite{Dornheim2}. Previously, the computational cost of $O(10^7)$ CPU hours could only simulate at most 34 fermions \cite{Bohme} under equal conditions \cite{Dornheim2}. Based on the extraordinary success of PIMC with fictitious identical particles for large-scale fermionic systems, Dornheim et al. \cite{Dornheim1,Dornheim2} strongly suggest that further research be conducted on the methods and applicability of fictitious identical particles. 

This work is based on the following two motivations:

1. The computational complexity of simulating fermionic systems in the fictitious identical particle PIMC framework is $O(N^2)$ \cite{Dornheim1,Dornheim2}. In the initial paper \cite{XiongFSP,Xiong-xi} on fictitious identical particle PIMD, we used the recursion formula for bosonic PIMD developed by Hirsheberg et al. \cite{HirshPIMD}, to simulate fermionic systems with a computational complexity of $O(PN^3)$. From the efficiency perspective, fictitious identical particle PIMD and the corresponding PIMC are still not competitive. Fortunately, Feldman and Hirshberg \cite{HirshImprove} have made significant improvements to the recursion formula in their latest study of bosonic PIMD, resulting in a computational complexity of $O(N^2+PN)$. It is clear that it is urgent and important to incorporate the latest recursion formula for bosonic PIMD into fictitious identical particle thermodynamics. This work will successfully achieve this and provide open source code to facilitate future researches. 
%In Feldman and Hirshberg's work \cite{HirshPIMD,HirshImprove}, the molecular dynamics part uses LAMMPS, while in our open source code, all PIMD and fictitious identical particle thermodynamics are integrated by C code.
% without the adoption of LAMMPS.

2. Despite being a simplified model, the numerical simulation of the two-dimensional Fermi-Hubbard model has been a long-standing unsolved problem \cite{LeBlanc,Qin}. In the two-dimensional Fermi-Hubbard model, exact numerical simulations of more than 100 fermions at any temperature involve significant computational cost.
%, and there have been no published reports of exact numerical simulations of more than 100 fermions at any temperature. 
In this paper, we will show how to use a small server to complete the simulation of more than 100 fermions in a two-dimensional periodic potential. Although we analyze a continuous-space model in fictitious identical particle PIMD, this analysis can clearly also promote the understanding of the simplified Fermi-Hubbard model. The current work also lays the foundation for the numerical simulation of the physics related to the three-dimensional Fermi-Hubbard model in the future, as well as the numerical simulation of large-scale strongly correlated fermion systems. One direct application of this simulation can be expected to be the experimental study of ultracold fermionic gases in optical lattices \cite{Fermilattice}.

The organization of this paper is as follows. In Sec.\ref{PIMC}, we introduce fictitious identical particle PIMD and the latest recursion technique that greatly improves computational efficiency. In particular, we give the virial estimator for energy. In Sec.\ref{xisec}, we briefly analyze the reasons for the success of fictitious identical particles in overcoming the fermion sign problem. In Sec.\ref{FH}, we introduce the two-component fermion model in a two-dimensional periodic potential. In Sec.\ref{mtemperature}, we use the isothermal extrapolation method to present the simulation results of the energy of fermions at medium temperatures. In Sec.\ref{ltemperature}, we use the constant energy semi-extrapolation method to present the simulation results of the energy of fermions at any temperature, and compare the energy results with those of the isothermal extrapolation. In Sec. \ref{more}, we discuss general strategies for reliable simulation of large-scale fermion systems. Finally, in Sec.\ref{SD}, we give a summary and discussion.

\section{The thermodynamics of fictitious identical particles with PIMD}
\label{PIMC}

For any quantum system, the partition function is
\begin{equation}
Z(\beta)=Tr(e^{-\beta\hat H}).
\end{equation}
Here $\beta=1/k_BT$ with $k_B$ the Boltzmann constant and $T$ the temperature. The parametrized partition function for single-component fictitious identical particles with a parameter $\xi$ can be written as
\begin{equation}
Z(\xi,\beta)\sim\sum_{p\in S_N}\xi^{N_p}\int d\textbf{r}_1d\textbf{r}_2\cdots d\textbf{r}_N\left<p\{\textbf{r}\}|e^{-\Delta\beta\hat H}\cdots e^{-\Delta\beta\hat H}|\{\textbf{r}\}\right>.
\label{Xipartition}
\end{equation}
We may also call the above parametrized partition function as partition function of fictitious identical particles. $S_N$ represents the set of $N!$ permutation operations. The factor $\xi^{N_p}$ is due to the exchange effect of fictitious identical particles. $\xi=+1$ for boson partition function, while $\xi=-1$  for fermion partition function. In addition, $\{\textbf{r}\}$ denotes $\{\textbf{r}_1,\cdots,\textbf{r}_N\}$. $N_p$ is a number defined to be the minimum number of times for which pairs of indices must be interchanged to recover the original order $\{\textbf{r}\}$ from $p\{\textbf{r}\}$. With the technique of path integral, the parametrized partition function $Z(\xi,\beta)$ can be mapped as a classical system of interacting ring polymers. It is this idea that eliminates the exponential complexity difficulty involved in numerically expressing the quantum many-body wave function.

A recursion formula \cite{HirshPIMD,HirshbergFermi} is found by Hirshberg et al., to calculate the partition function for both bosons ($\xi=1$) and fermions ($\xi=-1$). The computational complexity is $O(PN^3)$. Here $P$ denotes the number of beads for a single particle when path integral is used to express the partition function. Following this recursion formula, there are many researches based on PIMD for identical particles \cite{HirshbergFermi,Deuterium,Xiong4,Xiong2,Xiong5,Xiong6,Xiong7}. We emphasize that in a recent work by Feldman and Hirshberg \cite{HirshImprove}, a significant step is made for the recursion formula of identical bosons so that the computational complexity becomes $O(N^2+PN)$. After this improvement, the computational complexity of PIMD is at the same level as that of ordinary PIMC. It is clear that incorporating the latest techniques of Feldman and Hirshberg into fictitious identical particles is very valuable.

In this work, we incorporate the latest recursive method \cite{HirshImprove} for bosons into the recursive expression for the parameterized partition function. The specific results are as follows. 
\par
Firstly, based on the path integral formalism and the recursive method in Ref. \cite{HirshImprove}, the discretized partition function of Eq. (\ref{Xipartition}) is found to be:
\begin{equation}
Z(\xi,\beta)=\left(\frac{mP}{2\pi\hbar^2\beta}\right)^{PdN/2} \int e^{-\beta(V_\xi^{[1,N]}+\frac{1}{P}U)}d\mathbf R_1...d\mathbf R_N,
\label{partition}
\end{equation}
where $\mathbf R_i$ represents the collection of ring polymer coordinates $(\mathbf r_i^1,...,\mathbf r_i^P)$ corresponding to the $i$th particle. $P$ denotes the number of beads for a single particle. The system under consideration has $d$ spatial dimensions. $V_\xi^{[1,N]}$ considers the exchange effects of the fictitious identical particles with parameter $\xi$ by describing all the possible ring polymer configurations. $U$ is the interaction between different particles, which is given by
\begin{equation}
U = \sum_{l=1}^P V(\mathbf r_1^l,...,\mathbf r_N^l).
\end{equation}
Here $V$ denotes the interaction potential. The expression of $V_\xi^{[1,N]}$ is the key to consider the exchange effects of fictitious identical particles, which is given by the following recursive relation:
\begin{equation}
e^{-\beta V_\xi^{[1,N]}}=\frac{1}{N}\sum_{k=1}^N\xi^{k-1}e^{-\beta (E^{[N-k+1,N]}+V_\xi^{[1,N-k]})},
\end{equation}
where $E^{[N-k+1,N]}$ is defined as
\begin{equation}
E^{[N-k+1,N]}=\frac{1}{2}m\omega_P^2\sum_{l=N-k+1}^N\sum_{j=1}^P(\mathbf{r}_l^{j+1}-\mathbf{r}_l^j)^2,
\end{equation}
where $\mathbf{r}_l^{P+1}=\mathbf{r}_{l+1}^1$ unless $l=N$ where $\mathbf{r}_N^{P+1}=\mathbf{r}_{N-k+1}^1$. The harmonic strings connecting the ring polymers of different particles have the frequency $\omega_P=\frac{\sqrt P}{\beta\hbar}$.

\par
We note that evaluating all the $E^{[N-k+1,N]}$ directly would take $O(N^3P)$ computational time, since there are $O(N^2)$ of them in total and evaluating each one takes $O(NP)$ time. Fortunately, Feldman and Hirshberg \cite{HirshImprove} purposed an iterative scheme to evaluate all the $E^{[N-k+1,N]}$ by eliminating all the redundant calculations and the resulting computational complexity is only $O(N^2+NP)$.
\par
In order to carry out molecular dynamics simulations, we must also calculate the force resulting from the potential $V_\xi^{[1,N]}$, which is its gradient with respect to each bead coordinate $\nabla_{\mathbf{r}_l^j}V_\xi^{[1,N]}$. There are $O(NP)$ forces in total and the evaluation of each force takes $O(N^2)$ time from the recursive definition of $V_\xi^{[1,N]}$, so the total computational complexity is $O(N^3P)$ as well. Feldman and Hirshberg \cite{HirshImprove} also derived a formula for $\nabla_{\mathbf{r}_l^j}V_\xi^{[1,N]}$ that is mathematically equivalent to the direct definition but takes only $O(N^2+NP)$ time to calculate, we present the result as the direct application of this technique to fictitious identical particles as follows.
%For single-component fictitious identical particles, the recursive expression for the partition function is:
\par
Firstly, we define another set of potentials through

\begin{equation}
e^{-\beta V_\xi^{[u,N]}}=\sum_{l=u}^N\xi^{l-u}\frac{1}{l}e^{-\beta (E^{[u,l]}+V_\xi^{[l+1,N]})},
\end{equation}
from the new $V_\xi^{[u,N]}$ we define the connection probabilities $Pr[G[\sigma](l)=l']$ between particles $l$ and $l'$ as follows. For $l'\leq l$:
\begin{equation}
Pr[G[\sigma](l)=l']=\xi^{l-l'}\frac{1}{l}\frac{1}{e^{-\beta V_\xi^{[1,N]}}}e^{-\beta(V_\xi^{[1,l'-1]}+E^{[l',l]}+V_\xi^{[l+1,N]})}.
\end{equation}
For $l'=l+1$:
\begin{equation}
Pr[G[\sigma](l)=l+1]=1-\frac{1}{e^{-\beta V_\xi^{[1,N]}}}e^{-\beta(V_\xi^{[1,l]}+V_\xi^{[l+1,N]})}.
\end{equation}
For $l'>l+1$:
\begin{equation}
Pr[G[\sigma](l)=l']=0.
\end{equation}
Now we can calculate the gradients $\nabla_{\mathbf{r}_l^j}V_\xi^{[1,N]}$ efficiently. First, for "interior beads" (beads with indices $1<j<P$ for all $l$), the gradient is trivial:
\begin{equation}
\nabla_{\mathbf{r}_l^j}V_\xi^{[1,N]}=m\omega_P^2(2\mathbf{r}_l^j-\mathbf{r}_l^{j+1}-\mathbf{r}_l^{j-1}),
\end{equation}
while for "exterior beads", it has been shown that we can use the connection probabilities to calculate the gradients as
\begin{equation}
\nabla_{\mathbf{r}_l^1}V_\xi^{[1,N]}=\sum_{l'=1}^NPr[G[\sigma](l')=l]\cdot m\omega_P^2(2\mathbf{r}_l^1-\mathbf{r}_l^2-\mathbf{r}_{l'}^P),
\end{equation}
\begin{equation}
\nabla_{\mathbf{r}_l^P}V_\xi^{[1,N]}=\sum_{l'=1}^NPr[G[\sigma](l)=l']\cdot m\omega_P^2(2\mathbf{r}_l^P-\mathbf{r}_l^{P-1}-\mathbf{r}_{l'}^1).
\end{equation}
The overall complexity for the gradient calculation part is also $O(N^2+NP)$.
\par

The energy estimator is not given in the paper by Feldman and Hirshberg \cite{HirshImprove}. Here, we will give the result of the energy estimator for the fictitious identical particles, with the application of the virial theorem. Even for identical bosons with parameter $\xi=1$, we emphasize that the present virial estimator for energy is an important improvement, compared with the previous works \cite{HirshPIMD,HirshImprove,XiongFSP,Xiong-xi}, where the virial theorem is not used. 

Based on the partition function, the average energy is
\begin{equation}
E=-\frac{\partial\ln Z(\xi,\beta)}{\partial \beta}.
\end{equation}
It is well known that a direct energy estimator derived from the above relation has variance in the order of $O(\sqrt{P})$, so a different estimator known as virial estimator \cite{Herman} is devised which will make the variance as small as possible to minimize the error in the simulation of the energy.

When the periodic boundary condition is also considered, following the previous works\cite{Herman,Spada}, the virial estimator of energy in the context of fictitious particle PIMD is : 
\[
<E>=\frac{dN}{2\beta}+<\epsilon_\xi^{(N)}>+<\frac{1}{P}\sum_{j=1}^PV(\mathbf{r}_1^j,...,\mathbf{r}_N^j)>+<\frac{1}{2P}\sum_{j=1}^P\sum_{l=1}^N(\mathbf{r}_l^j-\mathbf{r}_l^1)\cdot \nabla_{\mathbf{r}_l^j}V(\mathbf{r}_1^j,...,\mathbf{r}_N^j)>,
\]
where 
\begin{equation}
\epsilon_\xi^{(N)}=-\frac{1}{2}m\omega_P^2[p(\mathbf{r}_1^1,...,\mathbf{r}_N^1)-(\mathbf{r}_1^P,...,\mathbf{r}_N^P)]\cdot[p(\mathbf{r}_1^1,...,\mathbf{r}_N^1)-(\mathbf{r}_1^1,...,\mathbf{r}_N^1)],
\end{equation}
averaged over all permutations $p$ for $N$ particles weighted by $\xi^{N_p}e^{-\beta E_p}$. Here $(\mathbf{r}_1^1,...,\mathbf{r}_N^1)$ and $(\mathbf{r}_1^P,...,\mathbf{r}_N^P)$ represent two dN-dimensional vectors, while "$\cdot$" denotes the inner product between two  dN-dimensional vectors. We note that for distinguishable particles ($\xi=0$), there is only one permutation with no particle interchanges and $\epsilon_{\xi=0}^{(N)}=0$. Following the original PIMD formulation \cite{HirshImprove}, a recursive relation for $\epsilon_\xi^{(N)}$ can be derived:
\[
\epsilon_\xi^{(N)}=\frac{\sum_{k=1}^N\xi^{k-1}(\epsilon_\xi^{(N-k)}-\tilde E_N^{(k)})e^{-\beta(V_\xi^{[1,N-k]}+E^{[N-k+1,N]})}}{\sum_{k=1}^N\xi^{k-1}e^{-\beta(V_\xi^{[1,N-k]}+E^{[N-k+1,N]})}},
\]
\[
\epsilon_\xi^{(0)}=0,
\]
\begin{equation}
\tilde E_N^{(k)}=\frac{1}{2}m\omega_P^2\sum_{l=N-k+1}^N(\mathbf{r}_l^{P+1}-\mathbf{r}_l^P)\cdot(\mathbf{r}_l^{P+1}-\mathbf{r}_l^1),
\end{equation}
where $\mathbf{r}_l^{P+1}=\mathbf{r}_{l+1}^1$ unless $l=N$ where $\mathbf{r}_N^{P+1}=\mathbf{r}_{N-k+1}^1$. 
\par
Similarly to the original bosonic PIMD algorithm, evaluating all $\tilde E_N^{(k)}$ takes $O(N^3)$ time (there are $O(N^2)$ of them and calculating each one takes $O(N)$ time). Fortunately, following the quadratic scaling bosonic PIMD algorithm purposed by Feldman and Hirshberg \cite{HirshImprove}, the computational complexity can be reduced to $O(N^2)$ using the following iterative equations:
\[
\tilde E_l^{(1)}=0,
\]
\[
\tilde E_l^{(u)}=\tilde E_l^{(u-1)}-\tilde E_o(\mathbf{r}_{l-u+2}^1,\mathbf{r}_l^P,\mathbf{r}_l^1)+\tilde E_o(\mathbf{r}_{l-u+2}^1,\mathbf{r}_{l-u+1}^P,\mathbf{r}_{l-u+1}^1)+\tilde E_o(\mathbf{r}_{l-u+1}^1,\mathbf{r}_l^P,\mathbf{r}_l^1),
\]
\begin{equation}
\tilde E_o(\mathbf{r}_1,\mathbf{r}_2,\mathbf{r}_3)=\frac{1}{2}m\omega_P^2(\mathbf{r}_1-\mathbf{r}_2)\cdot(\mathbf{r}_1-\mathbf{r}_3).
\end{equation}
It is worth noting that the virial estimator presented here is also suitable for systems without periodic boundary conditions, such as identical particles trapped in a harmonic potential.

We can also apply the new recursive techniques to the case of two-component fictitious identical particles.
Assume there are $N_\uparrow$ particles in state $|\uparrow>$ and $N_\downarrow$ particles in state $|\downarrow>$, the parametrized partition function for two-component fictitious identical particles is
\begin{equation}
Z(\xi,\beta)\sim\sum_{p_1\in S_{N_\uparrow}}\sum_{p_2\in S_{N_\downarrow}}\xi^{P_1}\xi^{P_2}\int d\textbf{x}_1d\textbf{x}_2\cdots d\textbf{x}_N\left<p_1p_2\{\textbf{x}\}|e^{-\Delta\beta\hat H}\cdots e^{-\Delta\beta\hat H}|\{\textbf{x}\}\right>.
\label{twocomponent}
\end{equation}
Here $N=N_\uparrow+N_\downarrow$. ${\textbf x}_j$ denotes both the spatial state and internal state $s_j$ ($|\uparrow>$ or $|\downarrow>$) of the $j$th particle. $p_1$ ($p_2$) is the permutation of particles with internal state $|\uparrow>$ ($|\downarrow>$). $P_1$ and $P_2$ are the permutation counts for permutations $p_1$ and $p_2$, respectively. 

The recursive formula for single-component fictitious identical particles can be directly extended to the case of two-component fictitious identical particles. The technical details of the implementation can be found in the open-source code. The reliability of the new algorithm and code has been verified by comparing the calculation results with those of the previous code. We also confirmed the consistency of the results, compared with Fig. 6 in Ref. \cite{Dornheim1} about the energy of spin-polarized electrons in a 2D harmonic trap simulated by PIMC for identical fictitious particles.

\section{General considerations on the use of fictitious identical particles for the simulations of fermion system}
\label{xisec}

\subsection{Fermion sign problem and fictitious identical particles}

For fermion systems, in PIMD/PIMC, for identical fermions of the same spin state, the energy of a fermion system at temperature $T$ can be generally simulated as follows. Due to the antisymmetry of the wave function with respect to the exchange of different fermions, based on Monte Carlo or molecular dynamics sampling:
\begin{equation}
E(T)=\frac{A(T)}{S(T)}.
\end{equation}
Here, $A$ and $S$ are both positive numbers. 

Due to the antisymmetry of the fermion wave function, here
\begin{equation}
S\sim e^{-N/T}.
\end{equation}
The average sign $S$ decreases exponentially with the increase of particles or the decrease of temperature $T$. Since the energy $E$ of a fermion system is finite, it must be that
\begin{equation}
A\sim e^{-N/T}.
\end{equation}

Unfortunately, for large-scale fermion systems or small-scale fermion systems at extremely low temperatures, $S$ and $A$ are too small to accurately calculate the energy of fermion systems using direct PIMD/PIMC. Numerical simulation becomes difficult \cite{Dornheim} when the average sign $S\sim 10^{-3}$, and it is unlikely to simulate meaningful results when $S< 10^{-4}$.

For fictitious identical particles, we set
\begin{equation}
E(\xi,T)=\frac{A(\xi,T)}{S(\xi,T)}.
\label{ExiT}
\end{equation}
In this case, we still have the following exponential behavior \cite{Dornheim1}
\begin{equation}
A\sim e^{\xi N/T},~~~S\sim e^{\xi N/T}.
\end{equation}
The exponential behavior of $A$ and $S$ can be explained by the expression of the partition function (\ref{Xipartition}), because the number of terms in the partition function increases exponentially with $N$ when $\xi\neq 0$. A rigorous mathematical proof of the exponential behavior of $A$ and $S$ is a mathematical problem worth studying in the future. When $\xi< 0$, we still face the difficulties caused by the sign problem. However, from a physical point of view, $\xi$ represents a type of exchange interaction of identical particles. When $\xi> 0$, it represents attractive exchange interaction, and the stronger the attractive exchange interaction becomes as $\xi$ increases. When $\xi< 0$, it represents repulsive exchange interaction, and the stronger the repulsive exchange interaction becomes as $\xi$ decreases. Therefore, for $\xi$, there is a monotonically decreasing relationship between $E(\xi,T)$ and $\xi$; while for $T$, there is a monotonically increasing relationship between $E(\xi,T)$ and $T$.

The simple and monotonic relationship of energy derived from physics above must be reflected in Eq. (\ref{ExiT}). However, this embodiment seems to be disturbed by the exponential behavior of $A$ and $S$. For example, we may set $A\sim e^{c_1\xi N/T}$  and $S\sim e^{c_2\xi N/T}$
more specifically. Here $c_1$ and $c_2$ are two possible different coefficients. However, once the coefficients $c_1$ and $c_2$ even have a slight difference, it will lead to completely unreasonable energy results when $N$ is large. If the energy is to be reasonable, then we must require $c_1$ and $c_2$ to be exactly equal when $N$ is large. The larger $N$ or the smaller $T$, the more precise $c_1$ and $c_2$ need to be equal. This consideration leads to the expression of energy can only be set to:
\begin{equation}
E(\xi,T)=\frac{f_1(\xi,T)g(\xi,T)}{f_2(\xi,T)g(\xi,T)}.
\end{equation}
Here, $g(\xi,T)$ represents the part that decreases exponentially with the increase of $N$ or the decrease of $\xi$, and decreases exponentially with the decrease of T. We do not need to pay attention to the specific expression of $g(\xi,T)$. $f_1(\xi,T)$ and $f_2(\xi,T)$ represent the non-exponential parts. As we have discussed from the physical point of view, the exponential behavior of the numerator and denominator must be canceled out, so we have:
 
\begin{equation}
E(\xi,T)=\frac{f_1(\xi,T)}{f_2(\xi,T)}.
\label{simple}
\end{equation}

\subsection{Isothermal extrapolation}

Since $f_1(\xi,T)$ and $f_2(\xi,T)$ in the above expression represent non-exponential behavior, they can be assumed to be relatively simple functions. Considering the monotonic behavior of $E(\xi,T)$ with respect to $\xi$, for the same temperature, we can consider setting $E(\xi,T)$ to:
\begin{equation}
E(\xi,T)\approx a+b\xi+c\xi^2.
\end{equation}
Based on the energy simulation results for $\xi\geq 0$, we can easily obtain the three coefficients $a,b,c$. Then, we have the energy of the Fermi system as $E(\xi=-1,T)=a-b+c$. Fortunately, the validity of this simple quadratic function can be judged in the exact numerical simulation for $0\leq \xi\leq 1$.

The above isothermal extrapolation method \cite{XiongFSP} has been shown to be highly efficient and accurate for simulating medium-scale and large-scale Fermi systems in the case of weak quantum degeneracy \cite{Dornheim1,Dornheim2}. A further improvement is the constant energy semi-extrapolation \cite{Xiong-xi}, which can be used to efficiently and accurately simulate the energy of fermions from zero temperature (high quantum degeneracy) to medium and high temperature (moderate and weak quantum degeneracy).

It is worth noting that in the above analysis, the exponential behavior of the sign factor caused by the fermion sign problem is turned from a difficulty into an advantage of the isothermal extrapolation with respect to $\xi$. Of course, this advantage also comes from the monotonic behavior of $E(\xi,T)$ with respect to $\xi$ and $T$ from the physical point of view. We also note that the above analysis is universally applicable due to the ubiquitous existence of the fermion sign problem, which means that the fictitious identical particles have application value in many important Fermi systems. One of the purposes of the current work is to discuss the application value in the strong correlated Fermi systems related to the Fermi-Hubbard model.

The above isothermal extrapolation can be applied in principle to density distribution, static structure factor, static linear density response function, and imaginary-time density correlation function, etc \cite{Dornheim1}. In the work of Dornheim et al. \cite{Dornheim2}, the isothermal extrapolation method and PIMC have been used to give highly accurate simulations of the static structure factor and static linear density response function of 1,000 electrons in the warm dense matter.

\subsection{Constant energy semi-extrapolation}

\begin{figure}[htbp]
\begin{center}
\includegraphics[scale=0.3]{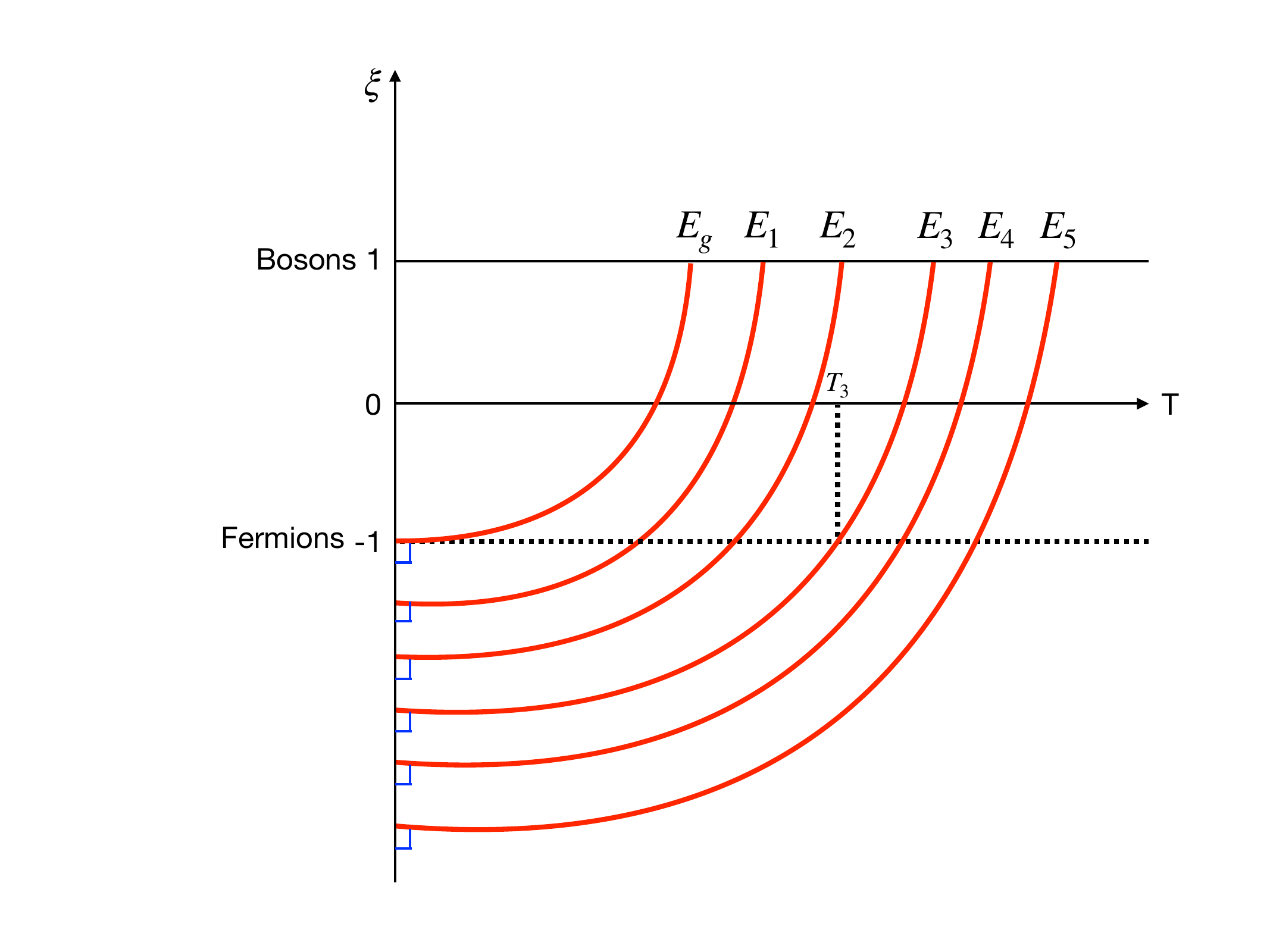}
\caption{The figure shows the basic method of constant energy semi-extrapolation.}
\label{equalE}
\end{center}
\end{figure}

In principle, the constant energy semi-extrapolation method is more accurate and has a wider range of applicability than the isothermal extrapolation method. One of the reasons is that in the constant energy semi-extrapolation, we can also use the following exact properties \cite{Xiong-xi}:
\begin{equation}
\left.\frac{\partial E(\xi,T)}{\partial T}\right|_{T=0}=0.
\label{confinement}
\end{equation}
In Fig. \ref{equalE}, we show how to use the constant energy semi-extrapolation to infer the energy of fermions. In the region where $\xi\geq 0$, we can obtain the constant energy curve $\xi_E(T)$ for a given energy $E$ and the simulated energy results. Compared with the isothermal curve $E_T(\xi)$, the constant energy curve $\xi_E(T)$ has one more condition (\ref{confinement}), so we think it is a semi-extrapolation method. In Fig. \ref{equalE}, we show different constant energy curves, where $E_g<E_1<E_2<E_3<E_4<E_5$. For the ground state energy of fermions, the constant energy curve corresponding to $E_g$  just passes through $\xi=-1$ and is perpendicular to the $\xi$ axis. For $E_3$, the curve should also be perpendicular to the $\xi$ axis. The intersection point of this constant energy curve and the horizontal line $\xi=-1$ determines the temperature at which the Fermi system has energy $E_3$.

Constant energy semi-extrapolation also provides the reasons and applicability of choosing the simple Eq. (\ref{simple}) in isothermal extrapolation to obtain the accurate energy of a Fermi system. Considering the dependence of the parameter $\xi$ on the exchange interaction in fictitious identical particles, there is a monotonically increasing relationship between $\xi$ and $T$ for all constant energy curves. Moreover, these constant energy curves $\xi_E(T)$ will not intersect. When the energy is relatively large, the constant energy curve is longer and has a smaller curvature, which makes it more linear in the region of $-1\leq \xi\leq 1$. This leads to the relationship between $E$ and $\xi$ being more linear when the temperature is given, which in turn makes the use of Eq. (\ref{simple}) in isothermal extrapolation reasonable. However, at low temperatures, especially $T=0$, isothermal extrapolation is likely to cause large deviations, because the curvature of the constant energy curve near $E_g$ may be large at this time. In this case, constant energy semi-extrapolation is a necessary method.
We note that when $E<E_g$, since the constant energy curve must always intersect the $\xi$-axis vertically at $T=0$, the curvature of these low-energy constant energy curves will increase as the energy decreases, resulting in the "black hole" region described in the paper \cite{Xiong-xi}. In the "black hole" region, the relationship between energy and $\xi$ may have a turning point in the region of $\xi\leq 0$ for a given temperature, which may lead to the failure of the isothermal extrapolation method. Once the constant energy curve corresponding to $E_g$ is outside the "black hole" region, the constant semi-extrapolation method can in principle obtain the accurate energy of a Fermi system at any temperature.

\section{Two-component Fermi systems in a two-dimensional periodic potential}
\label{FH}

\begin{figure}[htbp]
\begin{center}
\includegraphics[scale=0.8]{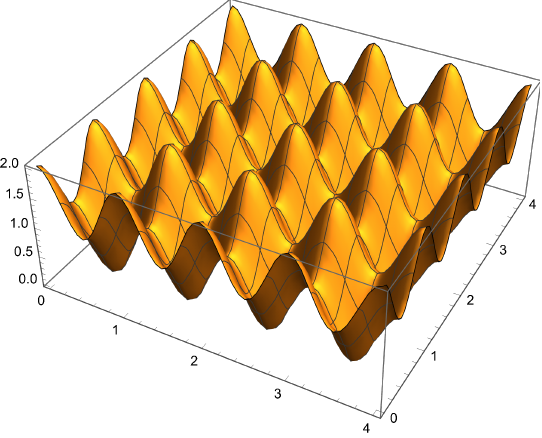}
\caption{
Shown is the two-dimensional periodic potential considered in this paper. More specifically, we also employ periodic boundary condition to study the lattice system with size $L\times L$. In the schematic diagram, the size of the lattice system is $5\times 5$, and the number of grids it contains is $5^2=25$.}
\label{potential}
\end{center}
\end{figure}

We consider $N_\uparrow$ and $N_\downarrow$  fermions in different spin states, respectively. The total number of fermions is $N_\uparrow+N_\downarrow$. We consider the fermions in the following two-dimensional periodic potential:
\begin{equation}
V(x,y)=h\left(\cos^2(\pi x)+\cos^2(\pi y)\right).
\end{equation}
We choose periodic boundary condition to carry out numerical simulations. The size of the box that satisfies the periodic boundary condition is $L\times L$ ($L$ is a natural number greater than zero), and the number of grids contained in the box is $L^2$ . We show the shape of this periodic potential in Fig. \ref{potential}.

For $N$ fermions, the Hamiltonian operator is:
\begin{equation}
\hat H=-\frac{1}{2}\sum_{l=1}^N\nabla_l^2+h\sum_{l=1}^N\left(\cos^2(\pi x_l)+\cos^2(\pi y_l)\right)+\sum_{l=1}^{N_\uparrow}\sum_{m=1}^{N_\downarrow}\frac{g}{\pi s^2}e^{-\frac{(\mathbf{r}^{\uparrow}_l-\mathbf{r}^{\downarrow}_m)^2}{s^2}}.
\label{periodic}
\end{equation}
Here $\mathbf{r}^{\uparrow}_l$ and $\mathbf{r}^{\downarrow}_m$ represent the coordinate of the fermion with spin $|\uparrow>$ and  $|\downarrow>$, respectively. We adopt the usual convention $\hbar=k_B=m=1$ in this paper.

In the numerical simulation of this paper, the parameters we choose are $h=1, g=1, s=0.5$. The value of $s$ we choose will make the interaction between particles mainly limited to the scope of a grid, in order to be as close to the Fermi-Hubbard model as possible; for this reason, we did not consider the interaction between fermions of different spin states in this paper. Of course, there is no problem in carrying out simulations after adding this interaction without extra difficulty.

\section{The energy results of isothermal extrapolation}
\label{mtemperature}

\begin{figure}[htbp]
\begin{center}
\includegraphics[scale=0.3]{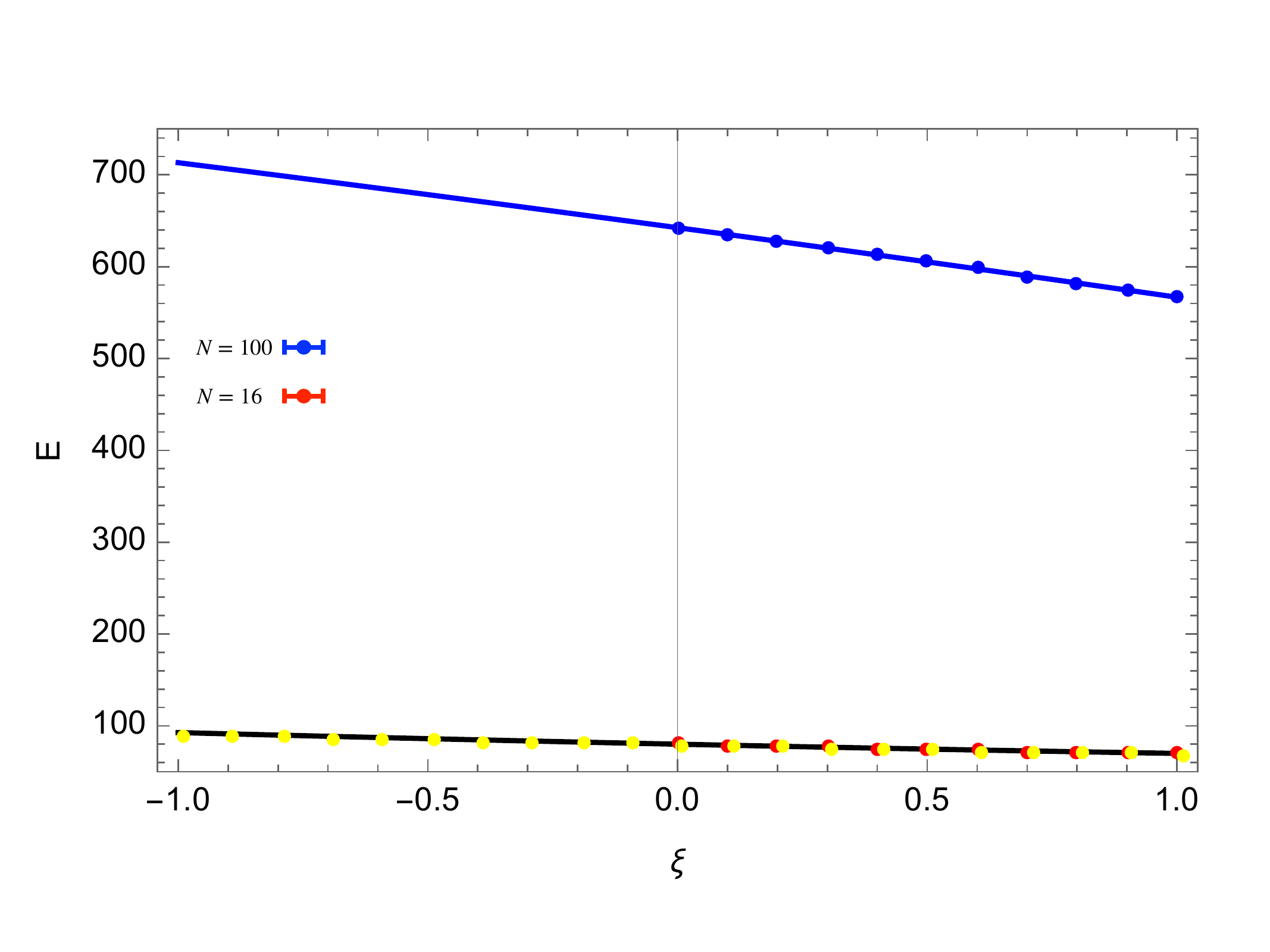}
\caption{The red points (PIMD) and yellow points (directly based on the expression of the canonical ensemble) are the calculated results of $16$ fermions in a $4\times 4$ box with periodic boundary condition. The black line represents the curve of isothermal extrapolation. The blue points are the energy of $100$ fermions in a $10\times 10$ lattice system under a two-dimensional periodic potential, based on the PIMD simulation. At the same temperature of $T=5$, the energy of different $\xi$ in the region of $\xi\geq 0$ is simulated by PIMD, and then the energy of $\xi=-1$, which is the energy of the fermions, is obtained by extrapolation. The blue line is the curve of isothermal extrapolation.}
\label{100fermions}
\end{center}
\end{figure}

To verify the rationality of the algorithm, we first calculate the energy of a two-dimensional periodic box. The Hamiltonian operator is as follows:
\begin{equation}
\hat H=-\frac{1}{2}\sum_{l=1}^{N_\uparrow} \nabla_l^2-\frac{1}{2}\sum_{l=1}^{N_\downarrow} \nabla_l^2.
\end{equation}
We consider a box of size $4\times 4$ with periodic boundary condition, and $N_\uparrow=N_\downarrow=8$. At the temperature $T=5$, based on the fictitious identical particle PIMD, we show the energy at different $\xi$ when $\xi\geq 0$ in Fig. \ref{100fermions} with red dots. The yellow dots are the energy of $-1\leq\xi\leq 1$ directly calculated by the expression of the canonical ensemble without using PIMD.
When directly calculating the energy in the canonical ensemble, we use the following recursive formula for the partition function of fictitious identical particles:
\begin{equation}
Z_\xi^{(N)}(\beta)=\frac{1}{N}\sum_{k=1}^N\xi^{k-1}z(k\beta)Z_\xi^{(N-k)}(\beta),
\end{equation}
\begin{equation}
z(\beta)=\sum_ie^{-\beta \epsilon_i}.
\end{equation}
$\epsilon_i$ is the single-particle energy level.
We note that the results obtained by the two different methods are consistent when $\xi\geq 0$, which proves the correctness of the PIMD algorithm and code.

At medium or high temperatures, we can accurately determine the energy of fermions through the method of isothermal extrapolation. Here we demonstrate this. In isothermal extrapolation, we use the quadratic function
$E(\xi)=a+b\xi+c\xi^2$.
After we determine the coefficients $a,b,c$ by fitting the region $\xi\geq 0$, the energy of the fermion is $a-b+c$. The black line in Fig. \ref{100fermions} is the curve of isothermal extrapolation. In the absence of interaction, we can calculate it without using PIMD under the canonical ensemble, and the result is represented by yellow points. We note that the yellow points agree with the entire black line. This result confirms our algorithm and code.

We now turn to the numerical simulation of the Hamiltonian operator (\ref{periodic}) with inter-particle interaction in the two-dimensional periodic potential. In the case of a $10\times 10$ lattice system, we simulated 100 fermions with $N=2N_\uparrow=100$ at a temperature of $T=5$. In the range of $0\leq \xi\leq 1$, we numerically simulated the energy for different $\xi$. The number of beads we used was $P=40$, and the MD steps we used in the simulation of massive Nos\'e-Hoover chain \cite{Nose1,Nose2,Hoover,Martyna,Jang} were $3\times10^6$. For the selection of the number of beads and MD steps, we need to check their convergence in actual research. In this case, the calculation of the energy for each parameter $\xi$ on a single CPU takes about 4 hours. For a multi-core computer, it is easy to break through the numerical simulation of more than 100 fermions by calculating for each $\xi$ simultaneously. In the past, even with the use of supercomputers, direct PIMD/PIMC for the Fermi-Hubbard model with hundreds of fermions was often hopeless.
In Fig. \ref{100fermions}, the blue points (PIMD simulation results) and the blue line (isothermal extrapolation curve) give the idea and results of isothermal extrapolation. As with the PIMC simulation of the energy of fictitious identical particles by Dornheim et al. \cite{Dornheim1}, we do not show the error bar in the figure because the fluctuation of the simulated energy can be neglected. We note that there are significant differences in the total energy between fermions, distinguishable particles, and bosons, indicating that quantum statistical effects or quantum degeneracy effects are very important in this example.
After fitting the quadratic polynomial about $E(\xi)$, we have $E(\xi)=642.46-73.25\xi-2.47\xi^2$. We note that the linear term plays a dominant role, so it can be expected that isothermal extrapolation can get a reasonable fermion energy.

For the half-filled case, the isothermal extrapolation method gives an average energy of 7.13 per fermion in the case of a $10\times 10$ lattice system. We also calculated the half-filled case of $8\times 8$ lattices (64 fermions) and $12\times 12$ lattices (144 fermions). We summarize the results in Table \ref{table:1}. We note that the average energy per fermion slowly decreases as the size of the box increases, as expected.

\begin{table}[h!]
\centering
\begin{tabular}{|c| c| c |} 
 \hline
 lattices & particle number &average energy per particle  \\ [0.5ex] 
 \hline
 $8\times 8$ & 64 & 7.27(6)  \\ 
 \hline
 $10\times 10$ & 100 & 7.13(6)  \\ 
 \hline
 $12\times 12$ & 144 & 7.02(6)  \\  [1ex] 
 \hline
\end{tabular}
\caption{Shown is the average energy per fermion for different numbers of lattice sites at $T=5$.}
\label{table:1}
\end{table}

\section{Constant energy semi-extrapolation and energy at any temperature}
\label{ltemperature}

The previous section (Sec.\ref{mtemperature}) analyzed the case of medium temperatures. For the cases of low temperature and zero temperature, we need to be especially careful. In the case of low temperature, when the energy is given to isothermal extrapolation, the real curve in the region of $\xi\leq 0$ may have a turning point, which may lead to systematic bias in the extrapolation. This problem can be solved by means of constant energy semi-extrapolation. In constant energy semi-extrapolation, we establish the constant energy curve $\xi_E(T)$ in the region of $\xi\geq 0$ for the given energy $E$, and then infer the temperature of fermions based on the setting of $\xi=-1$.
The reason why we think this is a semi-extrapolation method is that in determining the $\xi_E(T)$ curve, we not only use the simulation results of $\xi\geq 0$, but also use the exact relationship  
$\left.\frac{\partial E(\xi,T)}{\partial T}\right|_{T=0}=0$ to determine the $\xi_E(T)$ curve.
This constraint condition can play a key role in establishing the curve $\xi_E(T)$.

\begin{figure}[htbp]
\begin{center}
\includegraphics[scale=0.3]{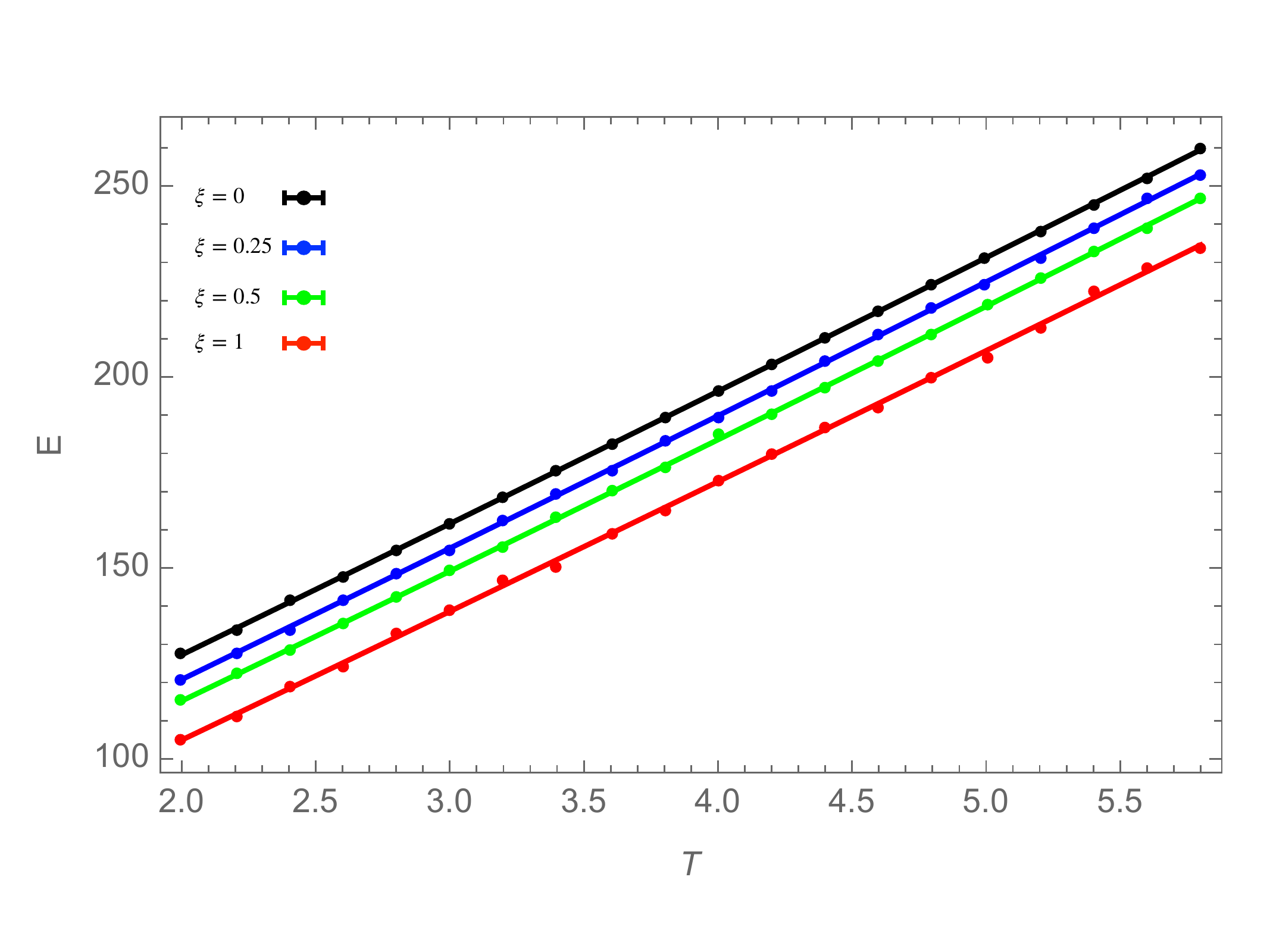}
\caption{The figure shows the energy simulation results of different $\xi$ at different temperatures in the case of $N_\uparrow=N_\downarrow=18$ in the two-dimensional periodic potential under a $6\times 6$ lattice system. From top to bottom, they are $\xi=0,0.25,0.5,1$. The solid lines are obtained by fitting with the quadratic function $E=a+bT+cT^2$.}
\label{36differentxi}
\end{center}
\end{figure}

To help readers understand the methods in this paper, we first analyze the case of a medium-sized $6\times 6$ lattice with half filling as an example. At this time, $N_\uparrow=N_\downarrow=18$. In the four cases of $\xi=0,0.25,0.5,1$, we calculate the energy in the temperature range of $2\leq T\leq 6$. In Fig.\ref{36differentxi}, we show the energy results of PIMD based on fictitious identical particles for different $\xi$ at different temperatures.

The constant energy semi-extrapolation method in this paper is improved over previous work \cite{Xiong-xi}, which has better numerical stability for the energy simulation fluctuations involved in $\xi\geq 0$. For a given energy $E$, we obtain the corresponding temperatures for different $\xi$ according to the results shown in Fig.\ref{36differentxi}, which means that we obtain four sets of numbers $\{\xi=0,T_1(E)\}$, $\{\xi=0.25,T_2(E)\}$, $\{\xi=0.5,T_3(E)\}$, $\{\xi=1,T_4(E)\}$. For these four sets of numbers, we use the following function to fit:
\begin{equation}
\xi+d(E)\xi^2=a(E)+b(E)T^2+c(E)T^3.
\end{equation}
After fitting to obtain the coefficients $a(E),b(E),c(E),d(E)$, we can determine the constant energy curve $\xi_E(T)$, and then infer the temperature $T(E)$ of the fermion system at the given energy $E$ after taking $\xi=-1$ and solving the equation above. Of course, in order to improve the fitting, we can also add different $\xi$ to carry out the calculation and verify the convergence.

\begin{figure}[htbp]
\begin{center}
\includegraphics[scale=0.3]{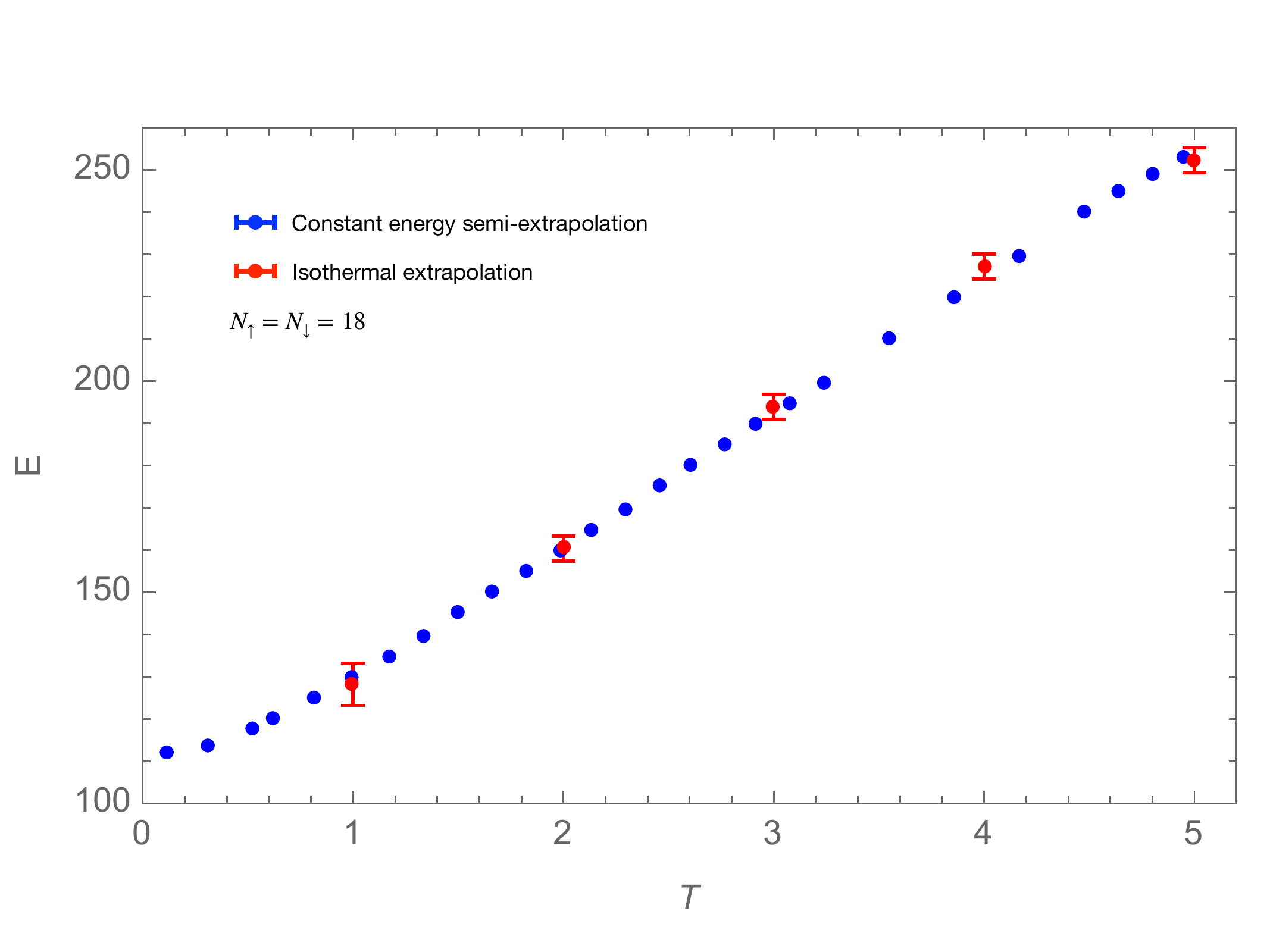}
\caption{The constant energy semi-extrapolation method for 36 fermions in a $6\times 6$ lattice system from low temperature to finite temperature (blue points). The red points in the figure are the results at temperatures $T=1,2,3,4,5$ under isothermal extrapolation. We note that when $T\geq 2$, even though the quantum degeneracy leads to a significant difference in energy between fermions and bosons, the results of isothermal extrapolation and constant energy semi-extrapolation are consistent. The constant energy semi-extrapolation is more accurate with smaller fluctuations. In the low temperature case of $T=1$, the fluctuations of isothermal extrapolation are very obvious, and the results of constant energy semi-extrapolation are more reasonable. The error bar of the red points shows the energy fluctuations at isothermal extrapolation.}
\label{36fermions}
\end{center}
\end{figure}

In Fig.\ref{36fermions}, we show the fermion energy from zero temperature to high temperature (blue points). For the case of medium and high temperature ($T\geq 2$), we also compare it with the isothermal extrapolation method (red points) and get consistent results. However, at low temperature, for example $T=1$, the energy given by isothermal extrapolation has larger energy fluctuations due to the relatively large curvature of the isothermal curve.
Compared with isothermal extrapolation, another advantage of constant energy semi-extrapolation is that it uses more simulation energy information of $\xi\geq 0$ when inferring the temperature for a given fermion system. Therefore, we notice that the fluctuations of the energy given by constant energy semi-extrapolation can be basically ignored.
Another advantage of constant energy semi-extrapolation is that it is much more efficient than isothermal extrapolation if we want to calculate the energy of fermion systems at all different temperatures. We also notice that in constant energy semi-extrapolation, the efficiency of simulating fermion systems at any temperature is even higher than that of boson systems, because for boson systems, we need to calculate the thermodynamic properties at any temperature separately.

\begin{figure}[htbp]
\begin{center}
\includegraphics[scale=0.3]{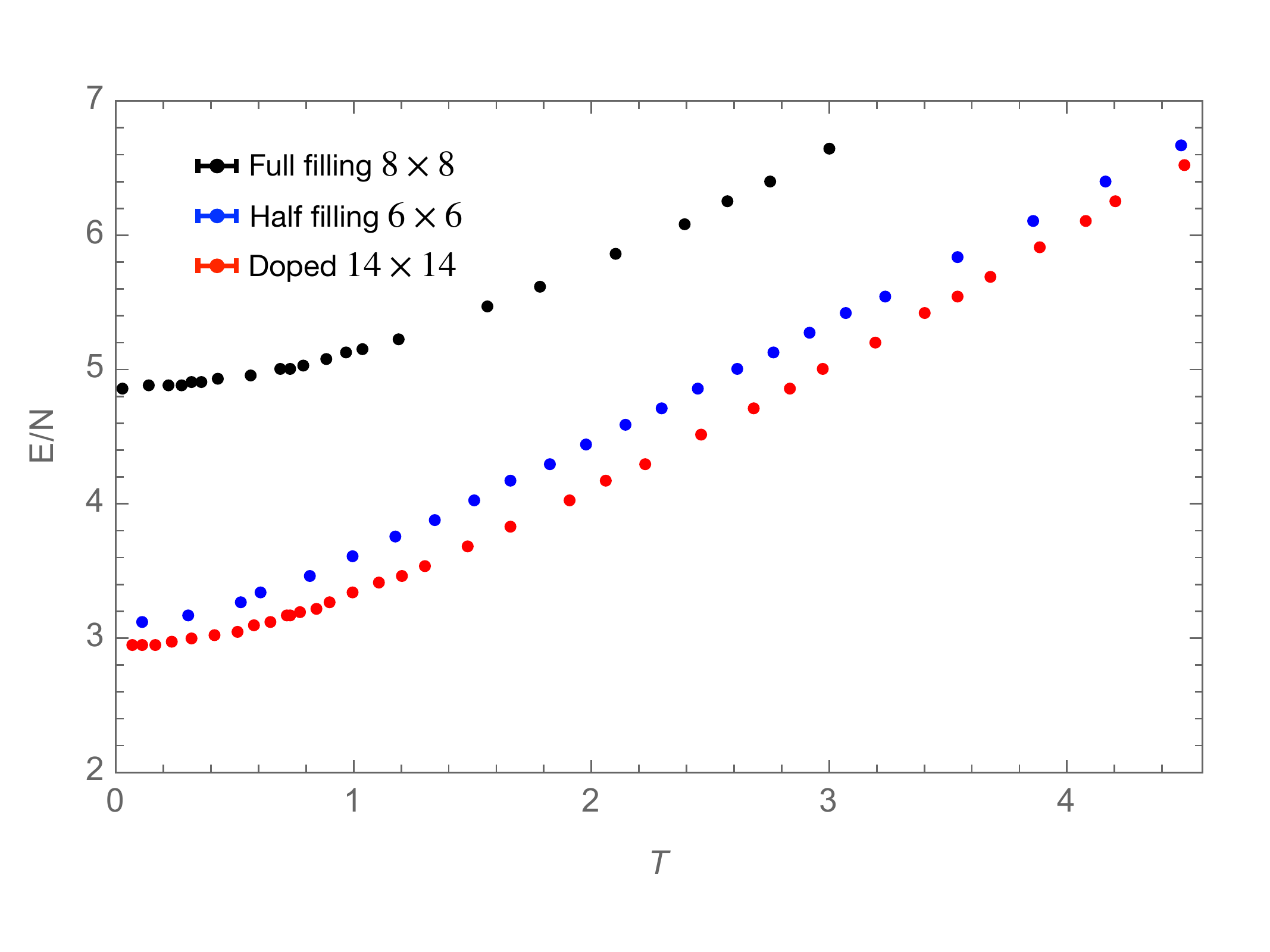}
\caption{The blue dots in the figure represent the average energy of each fermion in a $6\times 6$ lattice system at half filling. The red dots represent the average energy of each fermion in a $14\times 14$ lattice system with $N=2N_\uparrow=144$. The black dots represent the average of each fermion in a $8\times 8$ lattice system at full filling.}
\label{comparision}
\end{center}
\end{figure}

In conventional methods, the farther away from half-filled case, the more troublesome the problem of fermion sign problem becomes \cite{LeBlanc}. However, the calculation of the energy of a fermion system based on the thermodynamics of fictitious identical particles will not be disturbed by this factor. We will now calculate the case of $N=2N_\uparrow=144$ in a $14\times 14$ lattice system. In this case, it is equivalent to the case of doping in the Fermi-Hubbard model. In Fig.\ref{comparision}, we use red points to show the average energy of each fermion. The blue points are the average energy of each fermion in a $6\times 6$ lattice system under half filling. We note that the average energy of each fermion is higher at half filling, and the two tend to be consistent as the temperature tends to high, which agrees with physical expectations.

For the case of full filling, we can also simulate it. Here we simulate the case of $8\times 8$ lattices, $N=2N_\uparrow=128$ fermions, shown by black dots in Fig. \ref{comparision}. We see that the average energy of each fermion is significantly larger than the situation of half filling, which is reasonable from physical consideration.

\begin{figure}[htbp]
\begin{center}
\includegraphics[scale=0.3]{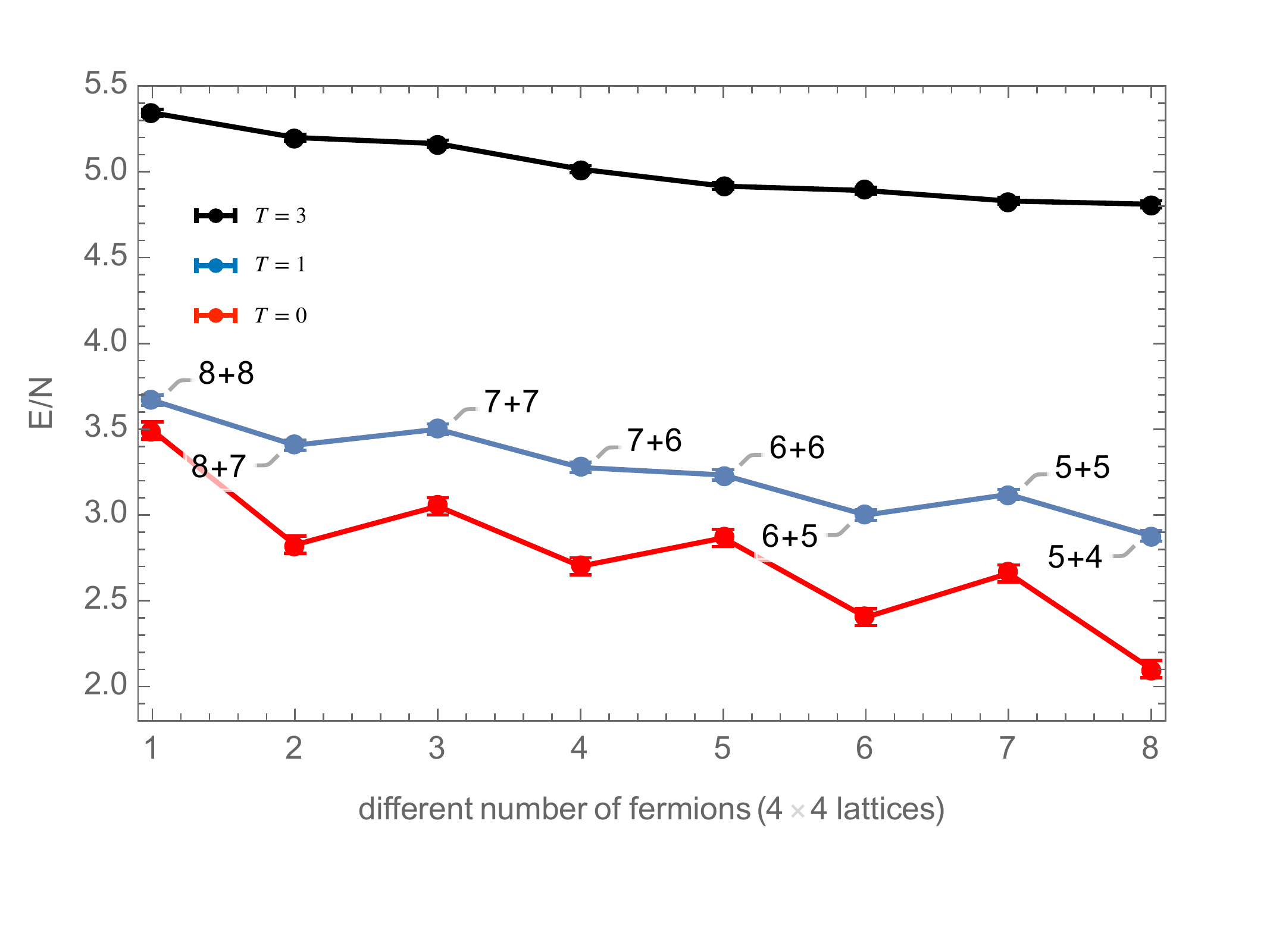}
\caption{
For a $4\times 4$ lattice system at different temperatures, we studied the average energy per fermion for $N_\uparrow+N_\downarrow=8+8, 8+7, 7+7, 7+6, 6+6, 6+5, 5+5, 5+4$. We note that at $T=0$, the ground state of the fermion system exhibits a clear sawtooth structure due to the antiferromagnetic effect. This sawtooth structure gradually disappears with increasing temperature.
}
\label{44lattices}
\end{center}
\end{figure}

As a final example, we use the constant-energy semi-extrapolation method to study the energy of 
different fermion numbers in a $4\times 4$ lattice. We start from the half-filled situation of 
$N_\uparrow=N_\downarrow=8$ and gradually reduce the number of fermions. In Fig.$\ref{44lattices}$, 
we show the average energy per fermion for different fermion numbers at different temperatures. We 
note that at $T=0$ (red dots), there is a sawtooth structure due to the antiferromagnetic distribution. 
When $N_\uparrow\neq N_\downarrow$, the antiferromagnetic effect is somewhat destroyed, leading 
to a lower average energy per fermion. This sawtooth structure gradually disappears with increasing 
temperature, as expected physically.

It is worth mentioning that in this paper, the medium temperature we mentioned refers to the case where the quantum degeneracy effect is significant (the energy of fermion systems and boson systems has a significant difference) but the isothermal extrapolation is applicable. Low temperature refers to the case where the isothermal extrapolation has obvious deviation and fluctuation, and thus requires constant energy semi-extrapolation.

\section{General strategies for reliable simulation of larger fermion systems}
\label{more}

From the physical and mathematical perspectives, if the thermodynamics of fictitious identical particles can handle dozens of fermions, then under the same physical conditions, we have the hope of handling thousands of fermions. We will discuss this under the condition of isothermal extrapolation, and these discussions are also suitable for constant energy semi-extrapolation.

Suppose that for a $L\times L$ lattice system with $N$ fermions, we can successfully infer the energy of the fermions through a quadratic function at temperature $T$, which is the following expression:
\begin{equation}
E(\xi, N)=a+b\xi+c\xi^2.
\end{equation}
Suppose that the above formula can accurately describe the case of $-1\leq\xi\leq 1$. Now we increase the fermion system by four times, that is, the number of lattices is four times, and the number of fermions is four times. Under the condition that other physical conditions have not changed, we naturally expect that the total energy will increase by about four times, that is, there is:
\begin{equation}
\frac{E(\xi, 4N)}{4}\approx a+b\xi+c\xi^2.
\label{4more}
\end{equation}
Therefore, we can assume that if the isothermal extrapolation can be established in the case of $N$ fermions in a $L\times L$ lattice system, then the isothermal extrapolation can also be established in the case of $4N$ fermions in a $2L\times 2L$ lattice system. Of course, the simple Eq. (\ref{4more}) does not accurately know the fermion energy of the larger system. We need to numerically simulate the energy of $4N$ fictitious identical particles in the case of $\xi\leq 0$. In this case, we will get
\begin{equation}
\frac{E(\xi, 4N)}{4}\approx \tilde a+\tilde b\xi+\tilde c\xi^2.
\end{equation}
Here, the coefficients  $\tilde a,\tilde b,\tilde c$ may have small differences from $a$, $b$, and $c$. Therefore, we still need to conduct numerical simulations on larger quantum systems to determine the new coefficients  $\tilde a,\tilde b,\tilde c$.

For large-scale fermion systems, it is difficult to have other numerical simulation methods for independent comparison. A reliable strategy is to start from a small system. For a small system of $L\times L$, we can use direct PIMD/PIMC or other methods to numerically simulate $E(\xi)$ in the entire range of $-1\leq\xi\leq 1$ and compare the results with the fictitious identical particle thermodynamics. If this comparison is successful, we can confidently use the fictitious identical particle method in a larger system of $2L\times 2L$ with the particle number increased to four times. By continuously increasing the scale of the system and monitoring the changes in the coefficients $a,b,c$, we can gradually simulate larger physical systems and assess their reliability. This strategy is also suitable for constant energy semi-extrapolation.

In their latest work, Feldman and Hirschberg \cite{HirshImprove} have successfully simulated over a thousand bosons using a new PIMD technique. Therefore, it is completely realistic to simulate thousands of fermions using fictitious identical particles and PIMD. This achievable goal will be an important collaborative approach for PIMC, which has already achieved the simulation of a thousand fermions using fictitious identical particles \cite{Dornheim2}.

\section{Summary and discussion}
\label{SD}

In summary, we have applied the latest recursive formula techniques of PIMD for bosons to PIMD for fictitious identical particles. We found that the computational efficiency for overcoming the fermion sign problem has been greatly improved compared to our previous work \cite{XiongFSP,Xiong-xi}. Our study shows that PIMD has the same important value as PIMC in simulating complex fermion systems. In addition, we applied the new algorithm of fictitious identical particle PIMD to a fermion system in a two-dimensional periodic potential, demonstrating that we have the potential to efficiently and accurately simulate the thermodynamic properties of fermions for this important physical system. In this paper, we considered continuous space fermion systems, but these studies still have the potential to promote the understanding of the Fermi-Hubbard model, which approximates actual systems. Additionally, how to directly apply fictitious identical particles to approximate lattice models such as the Fermi-Hubbard model is also a valuable research topic for the future.

The purpose of this paper is not to study the specific physical phenomena of the Fermi-Hubbard model, but to provide algorithms and codes, and to use the two-dimensional periodic potential as an example to illustrate our methods. It is a bit unfortunate that due to the great difficulties brought by the fermion sign problem, there are no other methods that can efficiently simulate hundreds of fermions in a periodic potential at any temperature. Therefore, we cannot compare our results with other theoretical results at this stage. However, based on various physical considerations, we have reason to believe that the methods and calculation results here are reliable, which provides a benchmark for the development of other methods to overcome the fermion sign problem \cite{nodes, Helium, Militzer, Mak, Blunt, Malone, Schoof1, Schoof2, Schoof3, Yilmaz, PB1, PB2, Joonho,Groth,SWZhang,QinMP,DornheimMod}, such as diagrammatic Monte Carlo \cite{diMonte,diMonte2,Burov2}.
One future scenario for testing the methods and applications is the periodic optical lattice of ultracold fermionic atoms. Based on the fictitious identical particle thermodynamics, PIMD/PIMC and the capabilities of supercomputers, it is realistic to simulate thousands of fermions from zero temperature to high temperature. Therefore, we have hope to compare directly with the experiments of ultracold fermionic atoms in the future \cite{Fermilattice,Bloch,Randeria,lattice}.

It is worth noting that in our previous work and this work, we mainly studied the energy of fermion systems based on fictitious identical particles. As pointed out by Dornheim's recent work \cite{Dornheim1,Dornheim2}, the $\xi$ extrapolation method can be applied to density distribution, static structure factor, static linear density response function, and imaginary-time density correlation function in addition to energy. How to make the $\xi$ extrapolation method more widely applicable to physical scenarios is also an important issue to be studied in the future.

As a mathematically constructed hypothetical particle, although the original purpose of fictitious identical particles was to overcome the fermion sign problem in order to simulate large-scale fermion systems, fictitious identical particles also have potential in the following two areas:

1. The physics involved in small negative $|\xi|$. Even without the $\xi$ extrapolation method, we can also use direct PIMD/PIMC to perform accurate numerical simulations when $\xi< 0$ and is close to 0. In this case, we can still observe the repulsive exchange force caused by $\xi<0$. Although this $\xi$ value cannot directly tell us the exact properties of the fermion system ($\xi=-1$), it may bring us important insights about fermion systems.

2. The duality between bosons and fermions. The $\xi$ parameter serves to continuously connect boson systems and fermion systems. For example, if we observe a phase transition of bosons, such as a specific heat peak, at $\xi=1$, then this specific heat peak may also be observed in a fermion system at $\xi=-1$. This duality between bosons and fermions may have potential application in the connection between superfluidity of bosons and superconductivity of fermions, and other phenomena.

\begin{figure}[htbp]
\begin{center}
\includegraphics[scale=0.5]{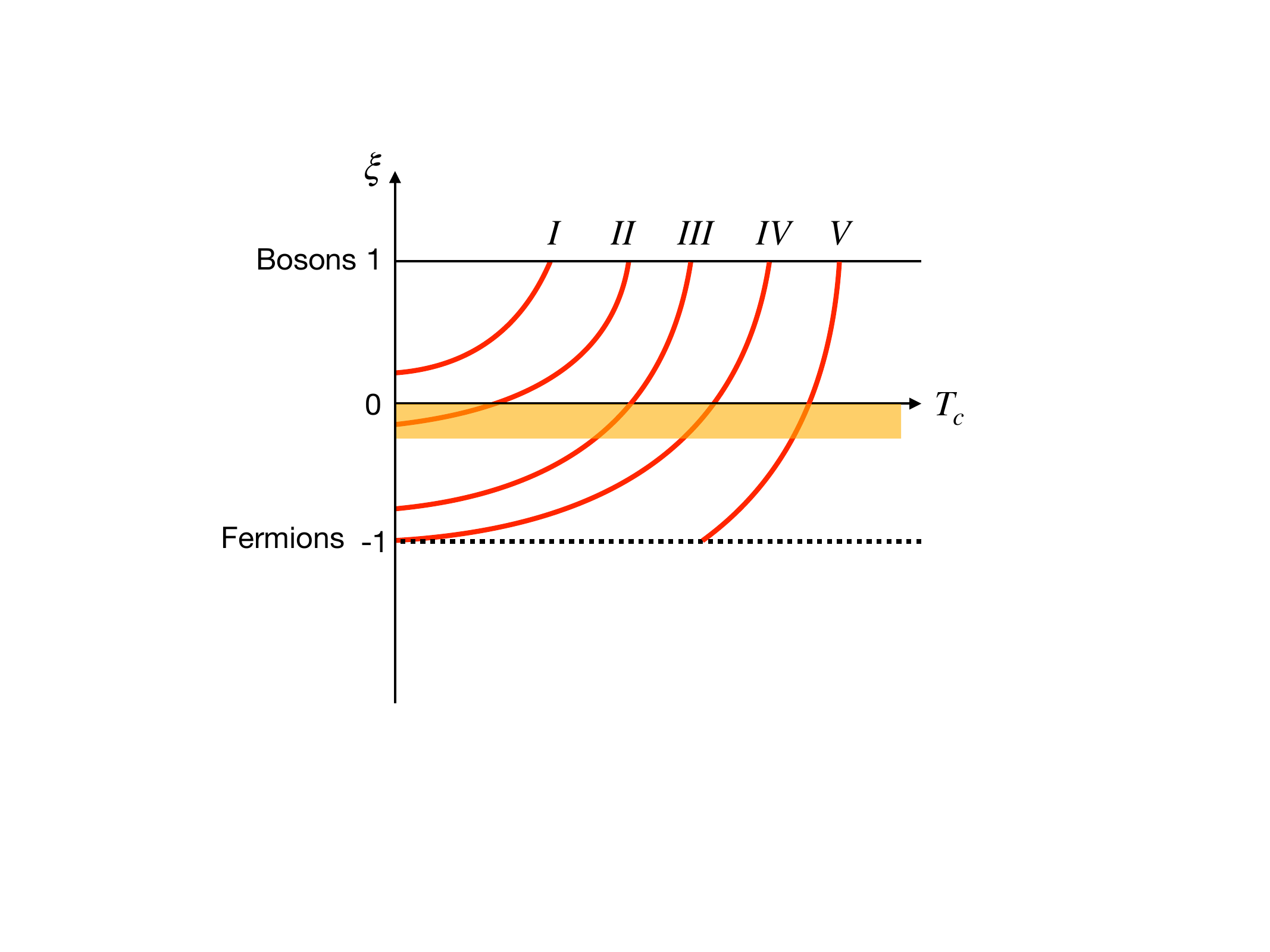}
\caption{Five possible phase transition situations shown in the $\xi-T_c$ plane.}
\label{phasetransition}
\end{center}
\end{figure}

In Fig.\ref{phasetransition}, we show five possible phase transition cases. In the figure, we show the curve resulting from the relationship between $\xi$ and the transition temperature $T_c$, that is, for each $\xi$, we assume that there is a corresponding transition temperature. For the first case, if we find that the phase transition curve $\xi(T_c)$ terminates at $\xi\geq 0$ in the region of $\xi\geq 0$, then we can determine that this phase transition curve will not reach the fermions. In Fig.\ref{phasetransition}, the light yellow area represents the region that PIMD/PIMC can directly simulate. Curve II represents the curve that terminates in the light yellow area. Curve III represents the termination at $\xi\geq -1$. Curve IV represents the termination at exactly $\xi=-1$. Curve V represents the phase transition occurring at a finite temperature in the fermion system, which is particularly interesting because it may have implications for low-temperature superconductor, high-temperature superconductivity and even room-temperature superconductivity.

The thermodynamics of fictitious identical particles provides a new dimension for analyzing important physical properties such as phase transitions. Considering that $\xi$ is a continuous real number, this new dimension has the potential to play a significant role. In this work, we mainly use PIMD (which is similar to PIMC) to analyze the thermodynamic properties of fictitious identical particles. Of course, we cannot ignore the significance of other important methods such as Hartree-Fock theory and density functional theory applied to $\xi$ physics. In any case, the thermodynamics of fictitious identical particles is a new field that is just beginning. It is beneficial to explore the hidden value in it by using various existing mature methods. The data that support the findings of this study are available from the corresponding author upon reasonable request. The code of this study is openly available in GitHub \cite{code}.

\begin{acknowledgments}
This work is partly supported by the National Natural Science Foundation of China under grant numbers 11175246, and 11334001. This work has received funding from Hubei Polytechnic University. We acknowledge the discussion with Prof. Xianqing Lin. 
\end{acknowledgments}

\end{document}